\documentclass[10pt,conference]{IEEEtran}
\IEEEoverridecommandlockouts\IEEEpubid{\makebox[\columnwidth]{ 978-1-6654-3540-6/22~\copyright~2022 IEEE \hfill} \hspace{\columnsep}\makebox[\columnwidth]{ }}
\usepackage{cite}
\usepackage{amsmath,amssymb,amsfonts}
\usepackage{graphicx}
\usepackage{textcomp}
\usepackage[dvipsnames]{xcolor}
\def\BibTeX{{\rm B\kern-.05em{\sc i\kern-.025em b}\kern-.08em
    T\kern-.1667em\lower.7ex\hbox{E}\kern-.125emX}}
    \usepackage{times}
\usepackage{lipsum}
\usepackage{mwe}
\usepackage{fancybox}
\usepackage{multicol}
\usepackage{color}
\usepackage{graphicx}
\usepackage{epsfig}
\usepackage{graphicx,latexsym}
\usepackage{amssymb,amsthm,amsmath}
\usepackage{array}
\usepackage{xy}
\usepackage{amsthm}
\usepackage{mathtools}
\usepackage{epstopdf}
\usepackage{algorithm}
\usepackage{algpseudocode}
\usepackage{algcompatible,lipsum}
\usepackage{cite}
\usepackage{ amssymb }
\usepackage{float}
\usepackage{multirow}

\usepackage{hhline}
\usepackage{url}
\usepackage{booktabs}
\usepackage{amsmath,algorithm,tabularx}
\usepackage{subcaption}
\usepackage{dblfloatfix}

\newcommand{\bl}{\textcolor{black}}

\DeclarePairedDelimiter\ceil{\lceil}{\rceil}

\renewcommand{\theenumi}{\arabic{enumi}}

\newcolumntype{C}[1]{>{\centering\arraybackslash$}p{#1}<{$}}
\makeatletter

\newcommand{\Rmnum}[1]{\expandafter\@slowromancap\romannumeral #1@}

\usepackage{mathtools,enumitem,lipsum}

\newcommand{\multiline}[1]{%
  \begin{tabularx}{\dimexpr\linewidth-\ALG@thistlm}[t]{@{}X@{}}
    #1
  \end{tabularx}
}

\newcommand{\algmargin}{\the\ALG@thistlm}   
\makeatother
\algnewcommand{\parState}[1]{\State%
    \parbox[t]{\dimexpr\linewidth-\algmargin}{\strut #1\strut}}
\makeatother

\makeatletter
\newcommand\fs@betterruled{%
  \def\@fs@cfont{\bfseries}\let\@fs@capt\floatc@ruled
  \def\@fs@pre{\vspace*{5pt}\hrule height.8pt depth0pt \kern2pt}%
  \def\@fs@post{\kern2pt\hrule\relax}%
  \def\@fs@mid{\kern2pt\hrule\kern2pt}%
  \let\@fs@iftopcapt\iftrue}
\floatstyle{betterruled}
\restylefloat{algorithm}
\makeatother

\begin{document}

\title{ Digital Twin-Driven Computing Resource Management for Vehicular Networks}

\author{\IEEEauthorblockN{Mushu Li\IEEEauthorrefmark{1},
Jie Gao\IEEEauthorrefmark{2}, Conghao Zhou\IEEEauthorrefmark{1},
Xuemin (Sherman) Shen\IEEEauthorrefmark{1}, and Weihua Zhuang\IEEEauthorrefmark{1}}
\IEEEauthorblockA{\IEEEauthorrefmark{1}Department of Electrical and Computer Engineering, University of Waterloo, Waterloo, ON, Canada\\
\IEEEauthorrefmark{2}School of Information Technology, Carleton University, Ottawa, Ontario, Canada\\
Email: \{m475li@uwaterloo.ca, jie.gao@ieee.org, c89zhou@uwaterloo.ca, sshen@uwaterloo.ca, wzhuang@uwaterloo.ca\}}}

\maketitle

\begin{abstract}

This paper presents a novel approach for computing resource management of edge servers in vehicular networks based on digital twins and artificial intelligence (AI). Specifically, we construct two-tier digital twins tailored for vehicular networks to capture networking-related features of vehicles and edge servers. By exploiting such features, we propose a two-stage computing resource allocation scheme. First, the central controller periodically generates reference policies for real-time computing resource allocation according to the network dynamics and service demands captured by digital twins of edge servers. Second, computing resources of the edge servers are allocated in real time to individual vehicles via low-complexity matching-based allocation that complies with the reference policies. By leveraging digital twins, the proposed scheme can adapt to dynamic service demands and vehicle mobility in a scalable manner. Simulation results demonstrate that the proposed digital twin-driven scheme enables the vehicular network to support more computing tasks than benchmark schemes. 
\end{abstract}


\section{Introduction}
To support cutting-edge vehicular applications, vehicular networks are evolving into the Internet of Vehicles (IoV). Integrating diverse network resources (e.g., communication, computing, sensing, and storage), IoV can perform compute-intensive tasks, such as objective detection, onboard augmented reality, and data fusion \cite{Zhuang}. While resource orchestration in highly dynamic vehicular networks is challenging, artificial intelligence (AI) can enable complex resource management and network control in IoV \cite{Xuemin}. 

AI-based resource management is anticipated to play a critical role in facilitating self-optimizing, adaptive, and low-cost next-generation networks. Many related approaches have been proposed to support computing resource allocation in IoV \cite{FMC, Anwar, Mushu1}. However, implementing AI-based resource management in vehicular networks remains a challenge. First, existing networks may lack sufficient computing and data resources to support extensive machine learning for network management. Second, the long training time of learning algorithms pose an obstacle to supporting delay-sensitive vehicular applications. 
Therefore, a novel network architecture is necessary to enable AI in vehicular networks.

Digital twin is a promising emerging paradigm for network architecture innovation \cite{Khan}. A digital twin is a digital representation of a physical network entity, e.g., a vehicle or a base station, which is synchronized with the physical network entity \cite{DT, Nguyen}. As virtualized network entities, digital twins can characterize features such as service demands and resource utilization status of a network, such as vehicular network. With digital twins, data collection and analysis can be tailored to individual vehicles and network infrastructure for efficient data collection and computing \cite{Sun, Liu,Liang2}. Additionally, while the status of vehicles differs (e.g., in vehicle speed, computing capability, etc.), digital twins can provide a basis for abstracting their characteristics and determining network policies, {thereby reducing the data size required in training machine learning algorithms and improving learning efficiency in AI-based resource management. }

In this paper, we investigate digital twin-driven resource management for computing offloading in vehicular networks. The objective is to minimize service requirement violation in terms of computing delay and service discontinuity caused by vehicle mobility. A central controller establishes two-tier digital twins to characterize vehicle status and service demands. The characteristics are used to find resource management policy references, referred to as policy maps, through a deep reinforcement learning (DRL) method. Following the policy maps, a matching-based approach allocates computing resources of multiple edge servers to individual computing tasks in real time.
The contributions of this work include: 1) a novel two-tier digital twin framework for vehicular networks to facilitate flexible AI-based computing resource management, 2) a scalable DRL-based resource management approach to capture network dynamics and computing demands via digital twins, and 3) a low-complexity matching-based approach to enable real-time computing resource allocation.

\section{System Model}
\subsection{Network Model}
\begin{figure}
    \centering
    \includegraphics[width=65mm]{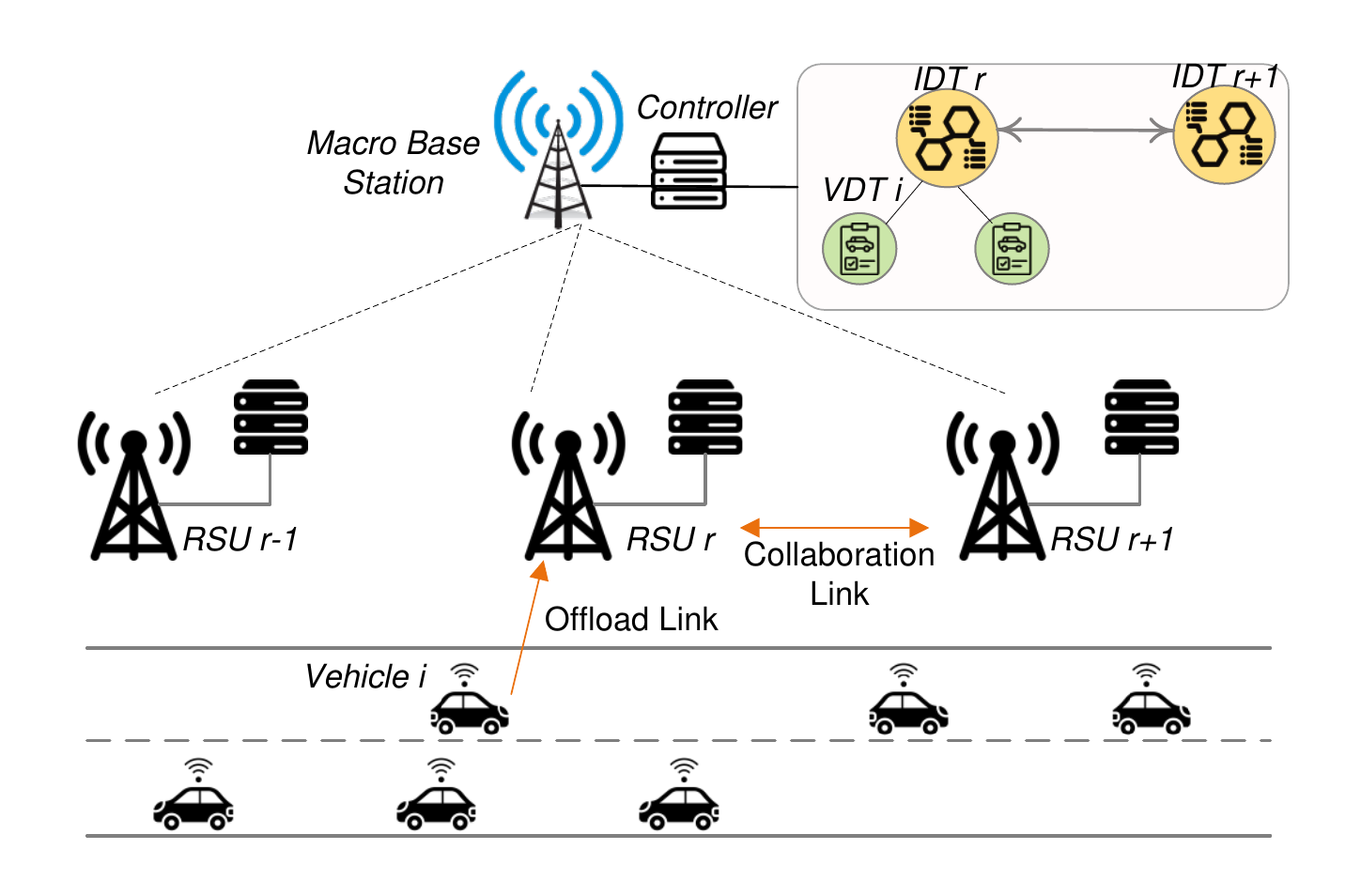}
        \vspace*{-2mm}
    \caption{An illustration of the considered network.}
    \label{fig:system}
    \vspace*{-3mm}
\end{figure}

{As shown in Fig. \ref{fig:system}, we consider a two-tier vehicular network on a one-dimensional road segment. The upper-tier consists of a macro-cell with a macro base station (MBS), and the lower-tier consists of $N$ micro-cells, each with one roadside unit (RSU). The MBS serves as a central controller to manage the RSUs resources by establishing computing policies.}
Edge servers located at the RSUs provide delay-constrained computing services (e.g., sensor data fusion for autonomous driving) to vehicles. Each vehicle  periodically generates computing tasks that require results within $\tau$ seconds from task generation. 

\setlength{\textfloatsep}{10pt}

{To enable scalable resource management, our computing resource allocation scheme involves two time scales. On a smaller time scale, the central controller allocates computing resources to each offloaded task in \textit{time slots}. The length of a time slot is constant and smaller than the delay threshold $\tau$. On a larger time scale, the network status is analyzed in \textit{policy epochs}, where policy epoch $k$ contains $K$ time slots starting from slot $t_k$, and $K$ is constant for all policy epochs.} 
Let $\mathcal{R} = \{1, \dots ,R\}$ denote the set of RSUs and the corresponding edge servers. In time slot $t$, the set of vehicles within the coverage area of RSU $r$ is denoted by $\mathcal{V}_{r,t}$, and the corresponding RSU connecting vehicle $i$ is denoted by $R_{i,t}$. Computing a vehicle's task at an edge server involves three steps: task offloading, task processing, and result delivery. 
\begin{itemize}[leftmargin=4mm]
\item Task offloading: Vehicle $i$ offloads its generated task to RSU $R_{i,t}$.  The communication rate for task offloading from a vehicle in $\mathcal{V}_{r,t}$ to RSU $r$ is constant over time (e.g., through transmit power control) and denoted by $h_{r}$. {We assume that all tasks to be offloaded have the same size, denoted by $H_I$.}
\item Task processing: The central controller selects an RSU to process the task, considering load balancing among RSUs. To minimize the communication overhead, we limit the edge server that processes the offloaded task from vehicle $i$ in time slot $t$ to be the one at RSU $R_{i,t}$ or either of its immediate neighbours. Denote the set of these three eligible RSUs by $\mathcal{N}_{i,t}$. In the case that the task is offloaded to $R_{i,t}$ but processed at a neighbour RSU, $R_{i,t}$ forwards the task to the neighbour RSU through a wired link. 
\item Result delivery: The central controller selects either RSU $R_{i,t}$ or the next RSU that vehicle $i$ is approaching, denoted by $R^+_{i,t}$, for computing result delivery. The selection is determined by whether vehicle $i$ will move out of the coverage of RSU $R_{i,t}$ during the computing session and is made in advance, i.e., at the instant of task offloading. 
\end{itemize}
The computing policy given by the central controller for vehicle $i$ to offload a computing task in time slot $t$ is denoted by $\mathcal{P}_{i,t} = \{P_{i,t}, D_{i,t}\}$, where $P_{i,t} \in \mathcal{N}_{i,t}$ and $D_{i,t} \in \{R_{i,t}, R^+_{i,t}\}$ represent the RSU processing the computing task and the RSU delivering computing result, respectively. 

\subsection{Digital Twin Model} \label{sec.dt}

To determine the computing policy, two types of digital twins, i.e., vehicle digital twins (VDT) and infrastructure digital twins (IDT), are created and utilized by the central controller. 
\subsubsection{Vehicle Digital Twin}
A VDT is a virtual replica of a vehicle in the network, which includes the following information about vehicle status:
\begin{itemize}[leftmargin=4mm]
\item Vehicle trajectory: 
Denote the location of vehicle $i$ in time slot $t$ by $x_{i,t}$. At any time, a VDT stores the locations of the vehicle in the current and the past $T_N-1$ time slots, i.e., $\{x_{i,t'}\}_{t' = t-T_N, \dots, t}$ in time slot $t$ .
\item Offloading events: A VDT records information about every task offloaded by a vehicle. We refer to the computing process of the task offloaded by vehicle $i$ in time slot $t$ as an \textit{event}, denoted by $e_{i,t}$. The offloading event captures: 1) the vehicle location when the task is offloaded, i.e., location $x_{i,t}$; 2) future vehicle speed during the maximum tolerable computing time $\tau$, denoted by ${v}_{i,t}$, which can be predicted from the vehicle trajectory (e.g., using neural networks (NNs) or ARIMA \cite{Liang});  and 3) the computing performance, including the computing delay, denoted by $d_{i,t}$, and service discontinuity (if any).  
\end{itemize}




In addition, the VDT can store routing-related information to facilitate mobility management. The locator/ID separation protocol (LISP) is used to deliver messages from RSUs to vehicles. Specifically, the routing-related information in a VDT is listed as follows:
\begin{itemize}[leftmargin=4mm]
\item Vehicle endpoint ID -- the IP address that uniquely identifies the vehicle within the network, which can also be used as the identity of the VDT. 
\item Current routing locator IP address -- the IP address to route the message towards the vehicle through the currently connected RSU, i.e. $R_{i,t}$.
\item Potential future routing locator IP address -- the IP address that will be used for routing computing results to vehicle $i$ through RSU $D_{i,t}$.
\end{itemize}
The central controller proactively manages mobility based on a vehicle's real-time location and routing locator information in the corresponding VDT\footnote{\bl{The information in VDTs is periodically collected and processed by RSUs and the central controller. The configurations of digital twins, such as collection frequency and data analysis methods, should be determined based on network dynamics.}}.


\subsubsection{Infrastructure Digital Twin}
IDTs capture the status of individual RSUs and the network environment, including:
\begin{itemize}[leftmargin=4mm]
\item User demand status: The central controller identifies the mobility and location profile, aggregated from data in VDTs, of vehicles within each RSU's coverage over a policy epoch.  {First, we divide the road segments covered by an RSU into $X$ levels and the vehicle speeds into $V$ levels. Then, we use a $(X\times V)$ matrix to represent the number of vehicles at discretized location $x$ with discretized future speed $v$ during maximum tolerable computing time $\tau$ in a policy epoch. The matrix is referred to as the vehicle status matrix, and the vehicle status matrix of RSU $r$ in policy epoch $k$ is denoted by $\textit{S}_{r,k}$.  }
\item {Resource provision} status: IDTs also record computing resource provision status at the RSUs, including the computing rates of the edge servers, and their queue lengths in each time slot. 
\end{itemize}
Through analyzing the IDT data, the central controller generates a reference of computing policies, referred to as a policy map, for each RSU. The policy map is then stored in the corresponding IDT for subsequent real-time decision making. 

\subsection{Computing Model}
We use the task of vehicle $i$ in time slot $t$ as an example to illustrate the computing procedure. Given a computing policy, $\mathcal{P}_{i,t}$, the computing delay consists of three parts: task offloading delay, task processing delay, and result delivery delay. {We assume that the time for determining the computing policy is much shorter than the computing delay and is negligible.}

The offloading delay is the time to offload the computing task of vehicle $i$ to RSU $P_{i,t}$, which may or may not be RSU $R_{i,t}$. 
Denote the data rate between two RSUs $r$ and $j$ by $W_{r,j}$, where $W_{r,r} = \infty$. Then, the task offloading delay for vehicle $i$ in time slot $t$ is:
\begin{equation}
    O_{i,t} =   H_I\left(\frac{1}{h_{r}} + \frac{1}{W_{r,j}}\right),  r=R_{i,t}, j =  P_{i,t}.
\end{equation}

Upon its arrival at RSU $P_{i,t}$, the task waits in a queue until the computing unit is available. We assume that the computing rate of one edge server can be different from that of others but is constant while processing different tasks. At RSU $r$, the computing rate is $C_r$. The processing delay, denoted by $U_{i,t}$, depends on the real-time queue length at RSU $P_{i,t}$, denoted by $Q_{i,t}$.

Once the task processing completes, the result should be delivered back to vehicle $i$. Denote the time slot when the task offloaded by vehicle $i$ in time slot $t$ is processed by {$Y(i,t)$}, defined as
\begin{equation}
    Y(i,t) = t+\ceil{\frac{O_{i,t}+U_{i,t}}{\epsilon}}  
\end{equation}
where $\epsilon$ is the duration of a time slot. Here, $Y(i,t)$ depends on the network status and resources allocated to individual tasks. The central controller examines whether the vehicle is covered by RSU $D_{i,t}$ in time slot $Y(i,t)$. 

{We assume that the process of establishing the link between RSU $P_{i,t}$ and $D_{i,t}$ is parallel to computing task processing, and the corresponding link establishment time (e.g., from creating potential future locator ID) is much shorter than the task processing time. If vehicle $i$ is not connected to $D_{i,t}$ in time slot $Y(i,t)$, additional time for result delivery is required: If ${R}_{i,Y(i,t)} \notin \{R_{i,t}, D_{i,t}\}$, service discontinuity occurs, which results in additional time to establish a new link for result delivery, denoted by $E_1$. Otherwise, i.e., ${R}_{i,Y(i,t)} = R_{i,t}$ and $R_{i,t}\neq D_{i,t}$, additional signalling delay, denoted by $E_2$, is necessary for configuring locator IDs. }
The overall result delivery delay is
\begin{equation}
    D_{i,t} =  H_O(\frac{1}{W_{j,r}} + \frac{1}{{h}_{r}}) + E_{i,t}, r = R_{i,{Y(i,t)}}, j = P_{i,{t}}, \label{eq.1}
\end{equation}
where $H_O$ is the data size of the computing results, and
\begin{align}
    E_{i,t} =& E_1 \cdot\mathbf{1}\{{R}_{i,Y(i,t)} \notin \{R_{i,t}, D_{i,t}\}\}\notag \\
    &+E_2 \cdot\mathbf{1}\{{R}_{i,Y(i,t)} = R_{i,t}, R_{i,t}\neq D_{i,t}\}
\end{align}
where function $\mathbf{1}\{X\}$ is 1 if condition $X$ is satisfied and 0 otherwise.

The overall delay for the task offloaded by vehicle $i$ in time slot $t$ is
\begin{equation}
F_{i,t} =     O_{i,t}+U_{i,t}+D_{i,t}.
\label{eq.scheduling_delay}\end{equation}
Service requirements are violated if the overall delay exceeds the tolerance, i.e., $F_{i,t}>\tau$, or if service discontinuity happens.

Next, we present a two-stage digital twin-driven resource management method for minimizing service requirement violations. In each policy epoch, the central controller creates a policy map, for each RSU. Then, in each time slot, real-time computing resource allocation and mobility management decisions for individual vehicles are made based on the policy maps {by the central controller}. 

\section{Digital Twin-Driven Resource Management}

\begin{figure}
    \centering
    \hspace*{-4mm}
    \includegraphics[width=80mm]{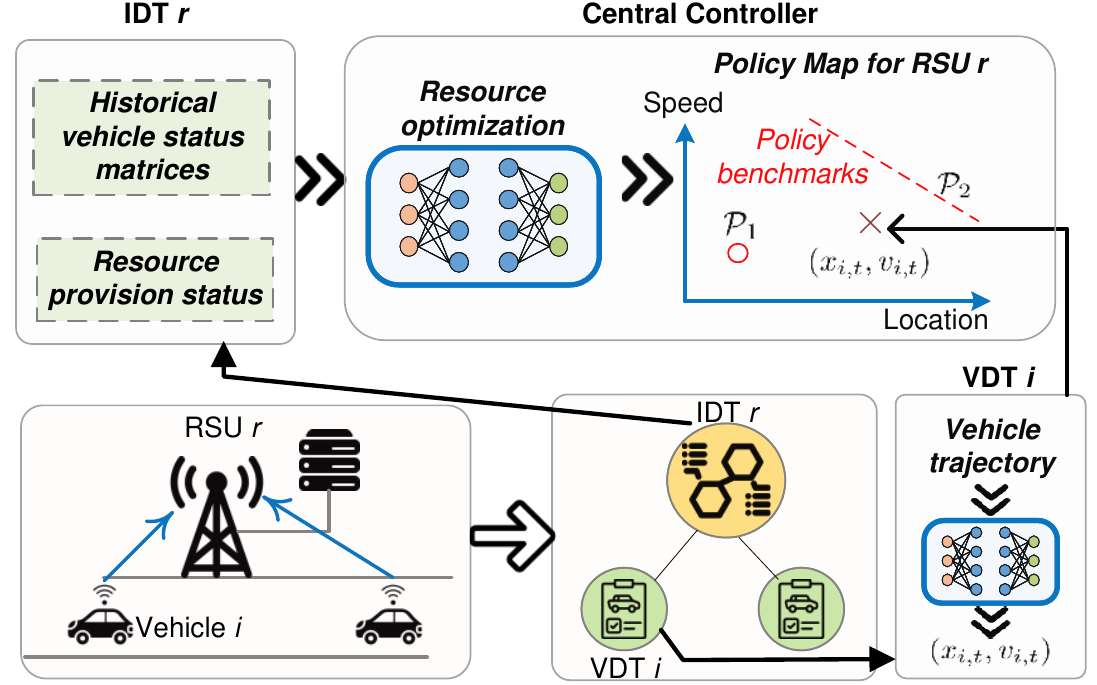}
    \caption{Digital twin-driven computing policy assignment.}
    \label{fig:dt}
    \vspace*{-3mm}
\end{figure}

In each policy epoch, the central controller generates a reference for making real-time computing policies for vehicles under the coverage of each RSU, which is referred to as a policy map. {The policy map is updated at the beginning of each policy epoch based on the resource provision status and the vehicle status matrices in previous policy epochs among RSUs.}
An example of a policy map for vehicles traveling on an one-way road is shown in Fig. \ref{fig:dt}.\footnote{{A policy map also applies to vehicles on two-way roads. }} {For each RSU $r$}, a policy map contains three elements: a 2-D coordinates on vehicle location and speed, computing policy benchmarks, and a reference queue length. 

\emph{Coordinates}: 
{The x-coordinate and y-coordinate of the policy map correspond to the two vehicle features stored in VDTs, i.e., location $x_{i,t}$ and future vehicle speed during the maximum tolerable computing time $v_{i,t}$, respectively.} The location and speed of a vehicle jointly determine the connectivity between a vehicle and the RSUs. {Specifically, if a vehicle is leaving the coverage of its connected RSU or driving away at a high speed, the next RSU is more likely to deliver the computing results. }

\emph{Computing Policy Benchmark}: A computing policy benchmark is a collection of points in the policy map. The benchmark depends on the network environment and is utilized to determine $D_{i,t}$ for vehicle $i\in \mathcal{V}_{r,t}$. \bl{There are two computing policy benchmarks for vehicles traveling in {each} direction.} The first policy benchmark (i.e., $\mathcal{P}_1$), as indicated by the marker ``o" in Fig. \ref{fig:dt}, is for when $D_{i,t} = R_{i,t}$. The second policy benchmark (i.e., $\mathcal{P}_2$), shown as a dashed line in Fig. \ref{fig:dt}, is for the case when $D_{i,t} = R_{i,t}^+$.\footnote{The format of computing policy benchmarks can vary depending on the network scenario and be configured.} 

\emph{Reference Queue Length}: The central controller generates a reference queue length, denoted by $\hat{Q}_{r,k}$, for the edge server in each RSU. {For load balancing, the queue lengths of edge servers should approach the reference queue lengths in real-time task offloading.}
The queue length parameter is determined in each policy epoch instead of each time slot to reduce the computing overhead for network management while capturing network dynamics.


After the policy maps are determined, in each time slot, the central controller makes decisions regarding mobility management and task offloading for individual vehicles based on their VDTs and the policy maps of their connected RSUs. Similar mobility management and computing policies apply to vehicles with similar locations and mobility. Therefore, the computing policy of vehicle can follow the closest benchmark, which is similar to {the features specified in its VDT}, in the 2-D policy map. Specifically, when vehicle $i$ generates a task for offloading in time slot $t$, {the central controller} checks the VDT and determines which RSU to deliver the computing result, i.e., the delivery RSU $D_{i,t}$, by solving the following problem:
\begin{align}
    {D}_{i,t} = \text{argmin}_{r\in \{R_{i,t}, R^+_{i,t}\}} d_r \label{eq.policy}
\end{align}
where 
\begin{align}
    d_r = \sqrt{(x_{i,t}-\hat{x}_{r,k})^2 +(v_{i,t}-\hat{v}_{r,k})^2}, \text{ if } r = R_{i,t}; \notag \\
    d_r = \frac{\|\alpha_{r,k}x_{i,t} + \beta_{r,k}v_{i,t} + \gamma_{r,k}\|}{\sqrt{\alpha_{r,k}^2 + \beta_{r,k}^2}}, \text{ if } r = R^+_{i,t}. \notag
\end{align}
The parameters of computing policy benchmarks $\{\hat{x}_{r,k}, \hat{v}_{r,k}, \alpha_{r,k},  \beta_{r,k}, \gamma_{r,k}\}$ are determined based on {historical vehicle status matrices} and resource provision information from the IDTs. Solving problem \eqref{eq.policy} is essentially about finding out which benchmark is closest to the target vehicle status $(x_{i,t}, v_{i,t})$.
After determining the delivery RSU $D_{i,t}$, the central controller determines $P_{i,t}$, i.e., the RSU to process the task. The corresponding policy is obtained through a matching-based approach, as detailed in Section~\ref{sec.match}.

Therefore, the computing resource management problem can be transformed into the problem of finding the parameters in the policy map, denoted by $\mathcal{L}_{r,k} = \{\hat{x}_{r,k}, \hat{v}_{r,k}, \alpha_{r,k},  \beta_{r,k}, \gamma_{r,k}, \hat{Q}_{r,k}\}$, for each RSU.
We aim to minimize the likelihood of delay requirement violation and service discontinuity. The optimization problem is 
\begin{align}
    \min_{\{\mathcal{L}_{r,k}, \forall r, k\}}& \frac{1}{T}\sum_{t = 1}^T {\frac{1}{\sum_{r\in \mathcal{R}}|\mathcal{V}_{r,t}|}}\sum_{{r\in \mathcal{R}, i\in \mathcal{V}_{r,t}}} I_{i,t}
    \label{eq.scheduling_opt}\\ 
    \text{s.t. }& I_{i,t} = \mathbf{1}\{F_{i,t}>\tau \} + w_{i} \mathbf{1}\{{R}_{i,Y(i,t)} \notin \{R_{i,t}, D_{i,t}\}\} \notag\\
    &\eqref{eq.scheduling_delay}, \eqref{eq.policy} \notag
\end{align}
where parameter $w_i$ weighs the significance of service discontinuity as compared to delay requirement violation for vehicle $i$, and $I_{i,t}$ measures the weighted service requirement violation. When $T \rightarrow \infty$, the problem is a long-term stochastic optimization problem. {Solving problem (7) can eliminate average service requirement violations as much as possible for tasks offloaded by vehicles in the considered road segment. } To capture service demand and network environment dynamics, we formulate the problem to a \bl{Markov decision process} (MDP) and obtain the policy maps through DRL in the next section. Based on the policy maps, a matching-based approach is proposed to allocate computing resources to individual vehicles in real time in Section~\ref{sec.match}.


\section{DRL-assisted Policy Map Establishment}
\bl{In this section, we propose a DRL-assisted algorithm to establish the policy map by solving problem \eqref{eq.scheduling_opt}.} We first formulate the stochastic optimization problem as an MDP. Define a tuple $(\mathcal{S},\mathcal{A}, \mathcal{T},\mathcal{C})$, where $\mathcal{S}$ represents the set of system states; $\mathcal{A}$ represents the set of actions; $\mathcal{T} $ is the set of transition probabilities; and $\mathcal{C}$ is the set of real-valued cost functions. The term $C(s,a)$ denotes the cost at state $s \in \mathcal{S}$ when action $a \in \mathcal{A}$ is taken. 
In the policy map establishment problem, the state space, action space, and cost model in the MDP are summarized as follows:
\begin{itemize}[leftmargin=4mm]
    \item State space -- The state in policy epoch $k$, denoted by $s_k$, includes the computing rate of RSUs, i.e., $\{C_r, \forall r\}$, the communication rate, i.e., $\{h_r, \forall r\}$, the current queue length at edge servers, i.e., $\{Q_{r,t_k}, \forall r\}$, the number of vehicles with computing tasks offloaded in each time slot between time slots $t-T_w$ and $t-1$ (i.e.,  $\{|\mathcal{V}_{r,t^*}|, \forall r,  t-T_w \leq t^* \leq t-1 \}$ ), and {the vehicle status matrices of all RSUs} during the previous policy epoch, i.e., $\{S_{r,k-1}, \forall r\}$. 
    \item Action space -- The action in each policy epoch is the parameters in the policy maps, i.e., $a_{k} = \{\mathcal{L}_{r,k}, \forall r\}$.
    \item Cost model -- The cost function can be formulated as
    \begin{align}
        C(s_k, a_k) = \frac{1}{K}\sum_{t = t_k}^{t_k+K} {\frac{1}{\sum_{r\in \mathcal{R}}|\mathcal{V}_{r,t}|}}\sum_{{r\in \mathcal{R}, i\in \mathcal{V}_{r,t}}}I_{i,t}.
    \end{align}
\end{itemize}
{The policy of the MDP, denoted by $\pi$,} is a mapping from $\mathcal{S}$ to $\mathcal{A}$, and we use a deep deterministic policy gradient (DDPG) method to obtain the policy.

\begin{algorithm}
\caption{DRL-based Computing Resource Management} 
\label{al.drl}
\begin{algorithmic}[1]
\STATE {Randomly initialize the parameters in NNs, i.e., $\{{\omega}, {\theta}, {\omega}', {\theta}'\}$.}
\STATE {Initialize random vector $\mathcal{N}_1 \sim (\varpi_1, \varsigma_1)$ as a noise for action exploration.}
\STATE {Initialize the replay memory. }
\STATE{Observe network state $s_1$.}
\FOR{policy epoch $k = 1: K$}
\parState{Select parameters $a_{k} = \mu(s_k|\boldsymbol{\theta})+\mathcal{N}_k$ in policy maps.}
\parState{Implement policy maps in each time slot by Alg. 2.}
\parState{Observe cost $c_k = C(s_k,a_k)$ and state $s_{k+1}$.}
\parState{Store tuple $(s_k, a_k, c_k, s_{k+1})$ to the replay memory. }
\State{Sample $N$ tuples from the replay memory.}
\STATE{Update $y_k$ by \eqref{eq.yt}. }
\STATE{Update weight ${\theta}_{k+1} = {\theta}_k - \phi \nabla J({\theta}_k)$.}
\STATE{Update weight $\boldsymbol{\omega}$ by minimizing the loss in \eqref{eq.omega}.}
\parState{Update target network parameters: $\theta' = \epsilon \theta' + (1-\epsilon) \theta$; $\omega' = \epsilon \omega' + (1-\epsilon) \omega$, where $\epsilon<1$.}
\STATE{Update noise vector $\mathcal{N}_{k+1} \sim (\rho \varpi_k, \rho \varsigma_k)$, where $\rho<1$.}
\ENDFOR
\end{algorithmic}
\end{algorithm}

Our DDPG method uses two types of deep NNs to learn the state value and policy at different speeds. The weights of \emph{evaluation networks} are updated in each training step, and the weights of \emph{target networks} are periodically replaced {by the weights} in evaluation networks. Both the evaluation and the target networks use an actor and a critic to evaluate the optimal policy and the Q value, respectively. For the actor in the evaluation network, the policy is parameterized by weights ${\theta}$, where $\pi(a|s,{\theta}) = P(a_k = a|s_k = s, {\theta})$. To learn ${\theta}$, the performance measurement function is defined and denoted by $J({\theta})$. The gradient descent method is applied to update ${\theta}$ by ${\theta}_{k+1} = {\theta}_k - \phi \nabla J({\theta}_k)$, where $\phi$ represents the learning rate of the actor. The gradient of $J(\boldsymbol{\theta})$ is  
\[\nabla J(\boldsymbol{\theta}_k) = \frac{1}{N} \sum_n  \nabla_a Q(s_k,\mu(s_k|\theta)|\omega)  \nabla_{\theta}\mu(s_k| \theta),
\]
where $\mu(s_k|\theta)$ represents the action taken at $s_k$ given weight $\theta$, and $Q(s_k,\mu(s_k)|\omega)$ represents the Q function approximated by the evaluation critic network with weight ${\omega}$. Parameter $N$ is the mini-batch size for training the networks, and $n$ is the index of the element in the mini-batch. In the evaluation critic network, the loss function $L(\omega)$ below is minimized:
\begin{equation}
    L(\omega) = \mathbb{E}{\big[( y_k - Q(s_k,\mu(s_k)|\omega) )^2\big]}. \label{eq.omega}
\end{equation}
The value of $y_k$ is approximated by the target network, where
\begin{equation}
    y_k = C (s_k, a_k) + \gamma Q(s_{k+1}, \mu'(s_{k+1}|\theta')|\omega'). \label{eq.yt}
\end{equation}
In \eqref{eq.yt}, $\mu'(s_{k+1}|\theta')$ represents the action taken at $s_{k+1}$ given by the target actor network with weight $\theta'$, $Q(s,a|\omega')$ represents the Q value for state-action pair $(s,a)$ given by the target actor network with weight $\omega'$, $\gamma$ is the learning rate of the evaluation critic network. The proposed DRL-based computing resource management scheme is shown in Algorithm \ref{al.drl}. 
\begin{algorithm}[t]
\caption{Matching-based Computing Resource Allocation} 
\label{al.matching}
\begin{algorithmic}[1]
\STATE{The central controller identifies $D_{i,t}$ via solving \eqref{eq.policy}.}
\STATE{VDTs and IDTs construct their preference lists, i.e., $\mathcal{E}^V_{i,t}$ and $\mathcal{E}^R_{r,t}$.}
\STATE{Each VDT proposes to its current most favorite RSU and then removes it from its preference list.}
\STATE{Each IDT checks all the received applicants, accepts its most preferred VDT proposals until the reference queue length $\hat{Q}_{r,k}$ is reached, and rejects the rest.}
\FOR{VDT $i$}
\STATE{If $\mathcal{E}^V_{i,t} = \emptyset$, $P_{i,t} = R_{i,t}$.}
\STATE{Otherwise, back to Step 3.}
\ENDFOR
\STATE{The algorithm is terminated when RSUs for processing all tasks offloaded in slot $t$ have been determined.}
\end{algorithmic}
\end{algorithm} 

\section{Matching-based Resource Allocation}\label{sec.match}

The central controller assigns computing policy $\mathcal{P}_{i,t} = \{P_{i,t}, D_{i,t}\}$ for vehicle $i$ in time slot $t$ following the policy maps {determined in Section IV}. Specifically, through the information provided by VDTs, problem \eqref{eq.policy} is solved to identify RSU $D_{i,t}$ for result delivery. Furthermore, to identify RSU $P_{i,t}$ for task processing, the central controller assigns computing resources of RSUs to vehicles following the reference queue lengths in the corresponding policy epoch, i.e., $\hat{Q}_{r,k}, \forall r$. 

We propose a matching-based approach to allocate computing resources to vehicles with the objective of minimizing computing delay, as presented in Alg. 2. When vehicle $i$ generates a computing task in slot $t$, the central controller checks its VDT and finds $D_{i,t}$ accordingly. Furthermore, its VDT constructs a preference list for selecting an edge server in RSU $r\in {\mathcal{N}}_{i,t}$ for task processing. The preference list is denoted by vector $\mathcal{E}^V_{i,t}$, which includes the indexes of RSUs in set ${\mathcal{N}}_{i,t}$ sorted by their current queue lengths, i.e., $Q_{r,t}, r\in {\mathcal{N}}_{i,t}$, in a non-decreasing order. A shorter queue means a shorter processing delay. Then, the IDT of RSU $r$ constructs a preference list for processing tasks, denoted by $\mathcal{E}^R_{r,t}$. The elements in $\mathcal{E}^R_{r,t}$ are in set $\{i|\forall r\in \mathcal{N}_{i,t}\}$ and sorted {based on offloading and result delivering delay.} {Assume that the input data size of a task is larger than the output data size.} For RSU $r$, offloading tasks from vehicles under its coverage, i.e., $i \in \mathcal{V}_{r,t}$, yields {the lowest offloading delay}. The corresponding vehicles have the highest ranks in the preference list. Next, offloading tasks from vehicles whose computing results will be delivered by it yields {a certain delay for offloading and the lowest delay for result delivery}. By contrast, offloading tasks from any other vehicle yields {the highest delay} compared with the above two cases. The idea of Alg. 2 is to assign computing tasks to edge servers in RSUs to comply with the reference queue lengths based {on the preference lists of both vehicles and RSUs}. 

\section{Simulation Results}
In this section, we conduct simulations to evaluate the performance of the proposed digital twin-driven resource management. Vehicles' trajectories follow taxi GPS traces in Didi Chuxing GAIA Initiative dataset. The simulation scenario includes 6 RSUs with edge servers and vehicles traveling in both directions on a 1.2 km road segment. Each time slot lasts for 0.5 seconds (s), and each policy epoch includes
10 time slots. Task input size $H_I$ is 0.2 Gbit. The signaling delay $E_1$ and $E_2$ are 1 s and 0.5 s, respectively. The delay requirement $\tau$ is 3 s. The weight $w_i$ in \eqref{eq.scheduling_opt} is 10. Four types of schemes are compared: \textit{DT+Matching} is the proposed scheme; \textit{DT only} is the proposed scheme without matching-based resource allocation; \textit{Migrate-x} delivers computing results to the next RSU when the vehicle is $x$ m or less before leaving the coverage of the currently connected RSU; and \textit{No-coop} only uses the connected RSU for offloading, task processing, and result delivery. \bl{For DDPG, we adopt a convolutional NN with a 3$\times$5$\times$3 convolution layer, a 2$\times$2 pooling layer, and a 3$\times$3$\times$3 convolution layer for both actor and critic, followed by fully-connected layers with [356,356,128,128] neurons for the actor and [128,64,64] neurons for the critic, respectively. }

\begin{figure}
    \centering
    \hspace*{-5mm}
    \includegraphics[width=101mm]{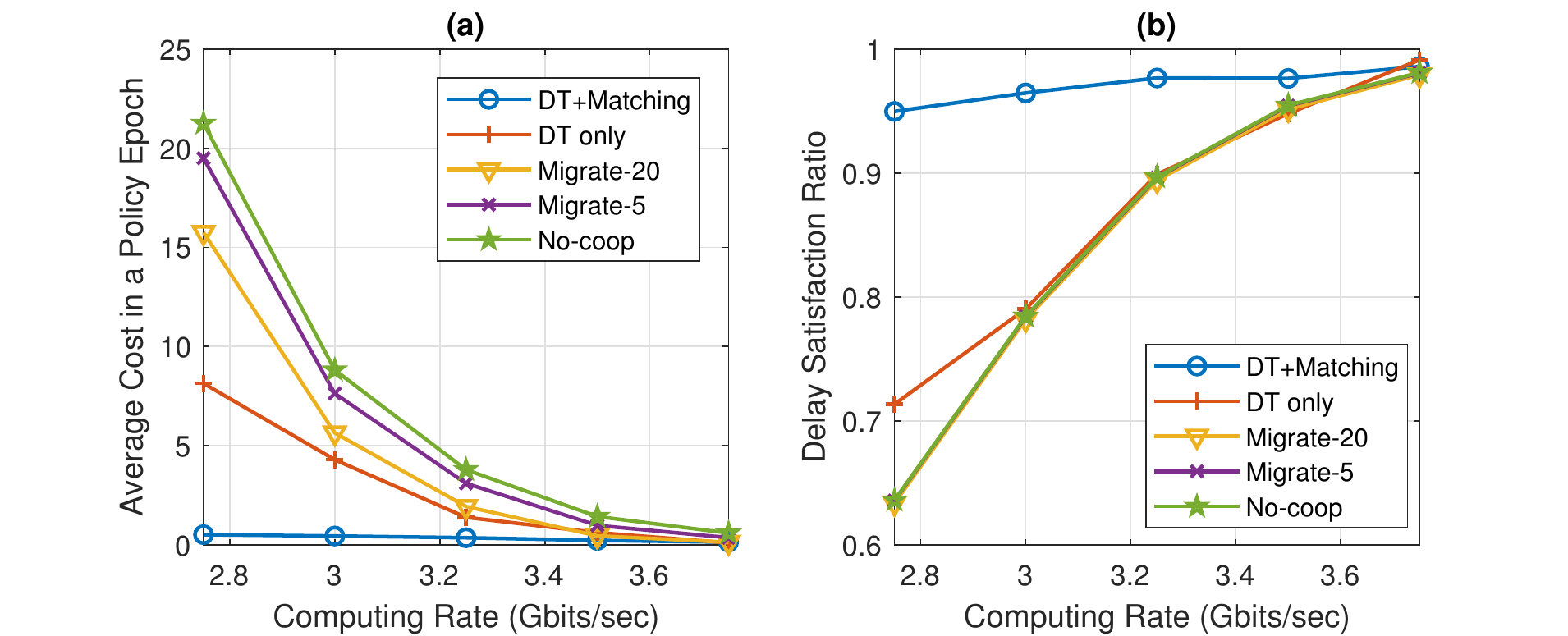}
    \vspace*{-2mm}
    \caption{(a) Average cost versus computing rate; (b) Average delay requirement satisfaction ratio versus computing rate.}
    \label{sim1}
    \vspace*{-3mm}
\end{figure}

Fig. \ref{sim1}(a) shows that the proposed scheme can significantly reduce the weighted service requirement violation compared to other benchmark schemes, especially when the computing rate is low. \bl{The reason is twofold: first, the proposed scheme balances computing loads among edge servers to minimize computing delay; second, the proposed DRL-assisted scheme identifies network dynamics via policy maps, such that the migration policy can be obtained for each task to avoid service discontinuity. Therefore, even if load balancing is not considered, i.e.,  the case of \textit{DT only}, using policy maps for decision-making still demonstrates outstanding performance.} Fig. \ref{sim1}(b) shows that the proposed scheme can maximize the delay requirement satisfaction ratio. The proposed scheme finds an optimal reference queue length that balances the loads among edge servers to minimize the overall computing delay.

\begin{figure}
    \centering
    \hspace*{-5mm}
    \includegraphics[width=101mm]{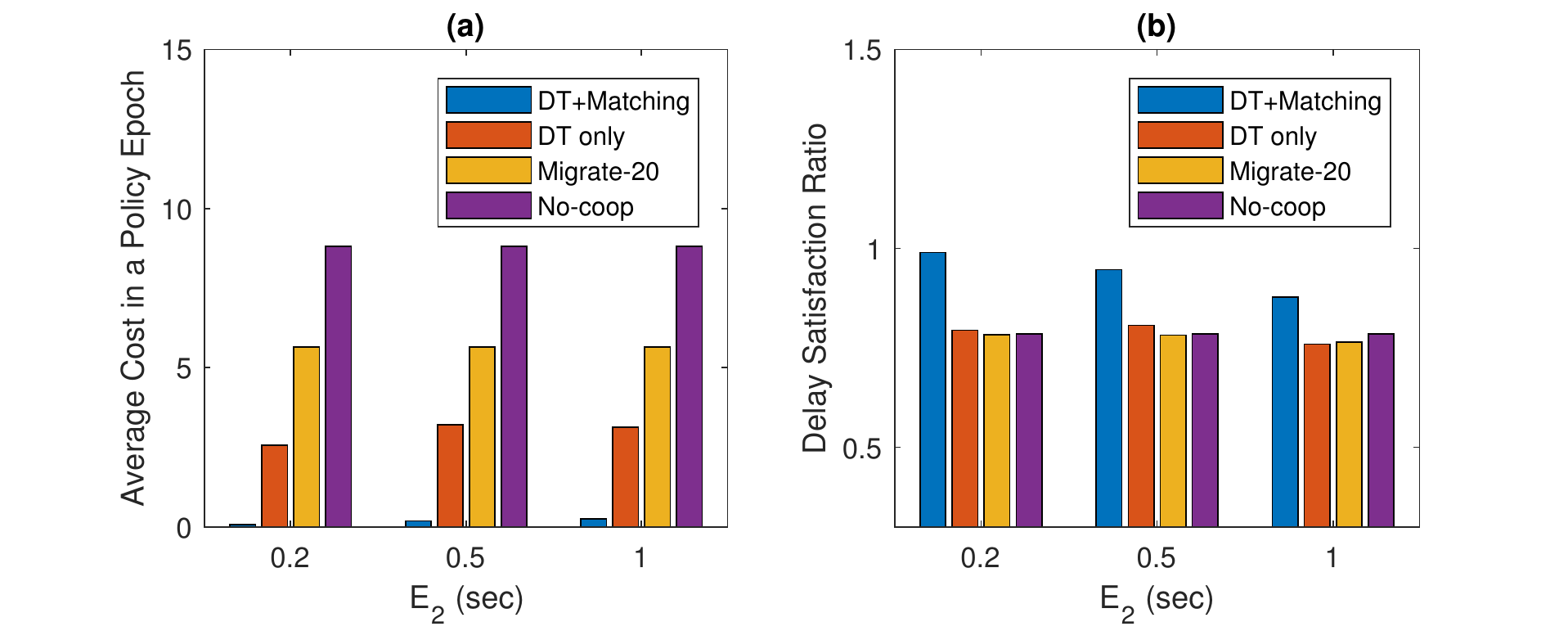}
    \vspace*{-2mm}
    \caption{(a) Cost in a policy epoch versus $E_2$; (b) Delay requirement satisfaction ratio versus $E_2$.}
    \label{sim3}
    \vspace*{-3mm}
\end{figure}
Fig. \ref{sim3} shows the performance of the proposed scheme with different {link reconfiguration time}, i.e., $E_2$. The cost of the proposed scheme increases with $E_2$ due to additional signaling overhead to migrate computing results in advance. For the same reason, the satisfaction ratio of delay requirements decreases as $E_2$ increases. However, the proposed scheme is able to find an optimal balance between such risk and potential performance gain. As a result, it always achieves the lowest cost or the highest delay satisfaction ratio as compared with the benchmark schemes.

\section{Conclusion}

In this paper, we have investigated digital twin-driven computing resource management for supporting delay-constrained computing tasks in vehicular networks. Leveraging digital twins, we have developed a novel machine learning-based resource management solution that uses policy maps to abstract the features of network environments and conducts matching-based resource allocation based on the policy maps, respectively. The proposed solution provides scalable learning-based resource management through an innovative digital twin architecture. The principle of this digital twin-driven approach holds promise for resource management in future highly dynamic networks. Our future work will further improve the efficiency of digital twin-driven resource management through the adaptive adjustment of policy epoch lengths.

\bibliographystyle{IEEEtran}
\bibliography{reference}

\end{document}